\pdfoutput=1
\documentclass[useAMS,usenatbib]{mnras}

\usepackage{graphics}
\usepackage{amssymb}
\usepackage{amsfonts}
\usepackage{wasysym}
\usepackage{epsfig}
\usepackage{breakurl}
\usepackage{comment}
\usepackage{longtable}
\usepackage{times}
\usepackage{mathptmx}
\usepackage[usenames,dvipsnames]{color}

\newcommand{\um}[1]{\,\mathrm{#1}}
\newcommand{\mic}[1]{\,\umu\mathrm{#1}}
\def\acen{$\alpha\,$Cen\,}
\def\deg{$^\mathrm{o} $\,}

\title[Radio Detection of $\alpha$ Centauri]
{Detection of $\alpha$ Centauri at radio wavelengths:\\ chromospheric emission and search for star-planet interaction}
\author[C.,~Trigilio et al.]
  {C.,~Trigilio$^1$\thanks{email: ctrigilio@oact.inaf.it},
  G.~Umana$^1$,  
  F.~Cavallaro$^{1,2}$,
 C. Agliozzo$^3$, 
  P.~Leto$^1$, 
\newauthor  
  C.S.~Buemi$^1$, 
  A.~Ingallinera$^{1}$,
  F. Bufano$^1$,
  S. Riggi$^1$,\\
  $^1$INAF- Osservatorio Astrofisico di Catania, Via S. Sofia 78,  95123, Catania, Italy\\
  $^2$Universit\`a di Catania, Dipartimento di Fisica e Astronomia, Via Santa Sofia, 64, 95123 Catania, Italy\\
  $^3$European Southern Observatory, Alonso de Cordova 3107, Vitacura, Santiago, Chile\\
  }
 \begin{document} 
\date{\textbf{}}

\pagerange{\pageref{firstpage}--\pageref{lastpage}} \pubyear{}

\maketitle
\label{firstpage}

\begin{abstract}
At radio wavelengths, solar-type stars emit thermal free-free and gyroresonance, gyrosynchrotron, and impulsive coherent emission. Thermal free-free emission originates at layers where the optical depth is close to unit, while high brightness temperature, variable emission, can be due to flares via gyrosynchrotron emission. We observed the \acen system with the Australian Telescope Compact Array at 2 GHz for three days and 17 GHz for one day. Both stars have been detected at 17 GHz, while only an upper limit has been obtained at low frequency despite the longer integration time. The brightness temperatures are consistent with the temperature of the upper chromosphere of the Sun. Inverting the formulae of the free-free emission, the average electron density of the plasma has been inferred. The same procedure was applied to the data in the millimetre recently acquired with ALMA. A comparison with the atmospheric solar models reveals a higher  level of activity in \acen B rather than in \acen A, even if still at quiescent level. The non detection at 2 GHz allows us to put a lower limit in the filling factor of active regions. The claimed detection of an Earth size planet in close orbit to \acen B, although doubtful, opens the opportunity to check the existence of Star-Planet Magnetic Interaction (SPMI). We constructed dynamic spectra in the 1.3 - 2.9 GHz of the 2 stars to search for time-variable coherent emission but obtained a null result.

\end{abstract}


\section{Introduction}

In the Sun and solar-type stars, 
characterized by a convective envelope and a radiative core, i.e. from late F to K main sequence stars of any age, radio emission arises mainly from three continuum emission mechanisms. They are thermal bremsstrahlung (or free-free), gyroresonance emission associated with active regions and non-thermal gyrosynchrotron, localized in the outer atmosphere at chromospheric and coronal levels. Thermal bremsstrahlung and gyroresonance emissions are not time variable on rapid timescales whereas gyrosynchrotron is normally associated with abrupt energy releases due to magnetic reconnection occurring during flares. It is also possible to observe coherent and impulsive bursts due to plasma emission and cyclotron maser.

Thermal bremsstrahlung is generated at a depth where $\tau_{\nu}\approx 1$. Since at radio wavelengths the opacity for free-free emission decreases with the frequency as $n_{\rm{e}}^2 T_\mathrm{e}^{-1.35} \nu^{-2.1}$, and  the plasma density $n_{\rm{e}}$ increases toward the surface, at low frequency we receive the radiation generated at high (coronal) levels, where the plasma temperature $T_\mathrm{e}$ is higher. Conversely, at high frequency, we see the chromosphere. Therefore with radio observations it is possible to probe the atmosphere, from the corona down to the chromosphere, without model assumptions, and get the density and temperature profiles \citep{white2004}. So far the quiescent photospheric or chromospheric continuum emission at centimetric wavelengths has been detected only in the supergiant Betelgeuse \citep{richards2013, lim1998} and the F5IV-V star Procyon \citep{drake1993}, which is evolving towards the red giant phase. No main sequence stars of the mass of the Sun has been detected at centimetric wavelengths. \citet{villadsen2014} observed three nearby main sequence solar-type stars, namely $\tau$\,Cet, $\eta$\,Cas\,A and 40\,Eri\,A with the JVLA, but at a higher frequency, 35\,GHz, where the photospheric emission is more intense since, as expected for a black body, it varies as $\nu^2$. In any case, the low S/N does not permit to use their detection to test atmospheric models.

Non-thermal emission occurs during flares in the Sun and solar-type stars. Sudden magnetic reconnections cause particle acceleration up to relativistic energies. Electrons eventually propagate in the magnetic structures, producing continuum gyrosynchrotron emission in the microwave band. 

In the Sun, coherent emission processes like Plasma Emission (PE) and Electron Cyclotron Maser (ECM), are often associated to flares and, in particular, to the acceleration processes (see e.g. \citealt{dulk1985, aschwanden2002} for a comprensive view). PE occurs at plasma frequency, $\nu_\mathrm{P}\approx 9000\,\sqrt{n_\mathrm{e}}\,$Hz, or its second harmonic, and is observed in type I and III bursts, where the decreasing frequency of the radiation reveals the decreasing density of the plasma for outward streams. Type II burst are associated with outward propagating shocks that can travel for several AUs. ECM occurs at gyrofrequency, $\nu_\mathrm{B}=2.8\times 10^6 B$\,(Hz), or its second harmonic, and is observed as short duration spikes (milliseconds) with a high degree of circular polarisation up to 100\% and very high brightness temperature ($T_\mathrm{B}>10^{13}$K) in a range of frequency from 100's MHz to several GHz. The ECM is observed in the Sun at the base of flaring loops.

Several types of  star  show intense coherent emission, like Cool {\citep{gudel2002} and Ultra Cool Dwarf stars (UCDs) \citep{hallinan2007}, close binaries as RS\,CVns and Algols \citep{white1995} and two magnetic CP stars \citep{trigilio2000, trigilio2008, trigilio2011, das2018}. There is also the suggestion that exoplanets could interact with the magnetosphere of the parent star \citep{hess2011, hallinan2015, leto2017}, similar to the Jupiter-Io interaction in the solar system, giving rise to coherent emission (see e.g. \citealt{zarka1998, zarka2004}). In this case, the frequency of the maser depends on the magnetic field intensity of the main body (parent star, Jupiter).

The quiet Sun is not a luminous radio emitter. In fact, \acen\,A, that can be considered a twin of our Sun, has never been detected so far at centimetric wavelengths, since its flux density should be of the order of few tens of $\mic{Jy}$ as it is proportional to $\nu^2$. It has been recently detected with ALMA at millimeter and sub-millimeter wavelengths \citep{liseau2015}, where the flux density is much higher.

\acen\ has is an ideal candidate for the study of the radio emission from solar-type stars. It is the closest stellar system to the Earth. The two main components of the system, \acen A and B, are in close orbit one with the other and can be observed simultaneously in the same field. The angular separation between them is about $4^{\prime\prime}$ at epoch 2015. In Table~\ref{param} we report the main characteristics of the two stars. The third component, Proxima Cen, is at angular separation of more than 2\deg from the main components.

\citet{dumusque2012} claimed the detection of a Earth size planet orbiting around \acen B (\acen Bb), detected with the method of the radial velocities (RV). \acen Bb was claimed to have an orbit of 0.04 AU from the parent star and a orbital period of 3.236 days. However, from a re-analysis of the RV data, \citet{Rajpaul2016} argued that the claimed detection of alpha Cen Bb was a false positive, since it could be due to a spurious ghost signal caused by the sampling of the data.

The presence of a planet could give us the opportunity to study any possible Star-Planet Magnetic Interaction (SPMI), that is important in the framework of the life development and protection in Earth-sized planets, even if the planet is not in the habitable zone. This information can be inferred by the existence, the type and the magnitude of the SPMI. When a planet crosses the magnetosphere of the parent star, the interaction depends on the cross area of the planet, including any planetary magnetosphere. Therefore, it is possible to argue the presence and the size of the magnetosphere. This is an important point since the magnetic field can protect the planet from the flux of charged particles generating during stellar fares and coronal mass ejections, allowing a quite environment for the development of the life. Furthermore, there is the possibility that SPMI could give rise to observable effects at radio wavelengths, since such an interaction can trigger ECM as in the case of the magnetized planets of the solar system.

With the aim to detect the thermal emission from chromosphere and low corona and any non-thermal emission due to flaring activity and ECM due to SPMI, we carried out radio observations of \acen with the Australia Telescope Compact Array (ATCA)\footnote{The Australia Telescope Compact Array is part of the Australia Telescope National Facility which is funded by the Australian Government for operation as a National Facility managed by CSIRO}. 
\\

\begin{table}
\caption{Parameters of the system $\alpha$ Centauri}
\label{param}
\begin{center}
\begin{tabular}{lcccc}
\hline\hline
					&	\acen A		&		\acen B		&			& ref \\
\hline
Spectral type			&	G2V 			&		K1V 			&		 	&      \\
Mass				&	1.13  		&		0.97 			& M$_{\odot}$ 	& (1) \\
Radius				&	1.2234		&		0.8632 		& R$_{\odot}$ 	& (2) \\
$L_\mathrm{bol}$		&	1.55			&		0.50 			& L$_{\odot}$	& (3) \\
$T_\mathrm{eff}$		&	5824			&		5223 		& K			& (3) \\
Angular diameter $\theta$	& 	8.502		&		5.999	 	& mas 		& (2) \\
parallax				& \multicolumn{2}{c}{743 $\pm$ 1.3 }			& mas	 	& (1) \\
\hline\hline
					&\multicolumn{2}{c}{Centre of Mass}			& 			& (4) \\
\hline
Coord (Eq 2000, Ep 2000)& 14:39:35.849 	& $-$60:50:07.63 		& 			&      \\
Proper motions			&		$-$3637 	&		694			& mas/yr	 	&      \\		
\hline
\hline
\end{tabular}
\end{center}
(1) \citet{Pourbaix2016}\\
(2) \citet{Kervella2017}\\
(3) \citet{Torres2010}\\
(4) \cite{Phillips}
\end{table}

\section{Observations and data reduction}
\label{observations}

The observations were carried out in January 2015, starting on Jan 22 and ending on Jan 26. In Table \ref{log-obs} the log of the observations is reported.
The observations were performed with the ATCA in the extended configuration (6-km) and the Compact Array Broadband Backend (CABB), with a bandwidth of 2 GHz for each band, in full polarization mode. For all bands, the flat-spectrum blazar 1253-055 (3C279) was observed as bandpass calibrator.
Similar observations were carried out in 2014 but discarded due to poor data quality.

At 2.1 GHz, the Seyfert-2 radiogalaxy 1934-638 was used to get the flux density scale (flux density F$_{2}$=12.6\,Jy). Phase corrections have been made by interpolating the data of the point-like radio-source 1414-59 (derived F$_{2}$=0.439 Jy), 2.9\deg far from the target. The duty-cycle was 2-20-2 minutes. Typical system temperatures during the observations in the three days were 40-55 K.

In $K$-band, 1934-638 (F$_{17}$=1.193 Jy) was observed as flux calibrator and the radio galaxy 0823-500 (F$_{17}$=0.585 Jy) was used to check the flux scale. We used as phase calibrator the source 1352-63 (derived F$_{17}$=1.244 Jy) 5.5\deg away, with a duty-cycle of 2-10-2 minutes. Typical $T_\mathrm{sys}$ during the observations were 70-100 K at 17 GHz and 220-250 at 23 GHz.

\begin{table}
\caption{Log of Observations}
\begin{center}
\begin{tabular}{ccccc}
\hline\hline
Day 			& Time 	&     Freq  		& Config	&	Time on source \\
 (UT)			&		&	(GHz)	&		&	(h:mm)	\\ 
\hline
2015 Jan 22-23	& 14:00-02:00	&	2.1		& A6 	&10:40 \\
2015 Jan 23-24	& 14:00-02:00	&	17, 23	& A6		& 08:10 \\
2015 Jan 24-25	& 14:00-02:00	&	2.1		& A6		&10:40 \\
2015 Jan 25-26	& 13:30-01:30	&	2.1		& A6		&10:00 \\
\hline\hline
\end{tabular}
\end{center}
\label{log-obs}
\end{table}

Data reduction and calibration were performed with the \texttt{MIRIAD} package \citep{Sault1995}. Data were firstly checked for RFI. Data for calibrators were edited interactively with the task \texttt{BLFLAG}, while data for \acen were automatically flagged for RFI by using the task \texttt{MIRFLAG}. The complex gains were determined with the standard \texttt{MIRIAD} procedures. Calibrated visibilities were exported and the mapping process performed with the Common Astronomy Software Applications (\texttt{CASA}) package \citep{McMullin2007}.

\section{Results}
\label{results}

\acen is the closest stellar system to the Earth, consisting of two solar-type stars A and B in a mutual orbit and the star Proxima Cen, about 12000 AU far away. The orbital period of \acen A and B is 79.91 yr, the orbit has an eccentricity of 0.5179, and the major axis $17.66^{\prime\prime}$. The orbital parameters have been recently accurately determined by \citet{Pourbaix2016}. Due to the proximity, proper motions are very high (Table\,\ref{param}), and therefore it is very important to get accurate positions of the two stars at the time of the observations. 

We computed the actual coordinates taking into account proper motion and parallax in order to determine the position of the barycenter of the system, then the angular distance and position angle of the vector connecting A and B projected in the plane of the sky and, finally, the dislocation of A and B relatively to the centre of mass based on their mass ratio. 
For this purpose, we used the procedure by \citet{Phillips}.

\begin{figure}
\includegraphics[width=8.5cm]{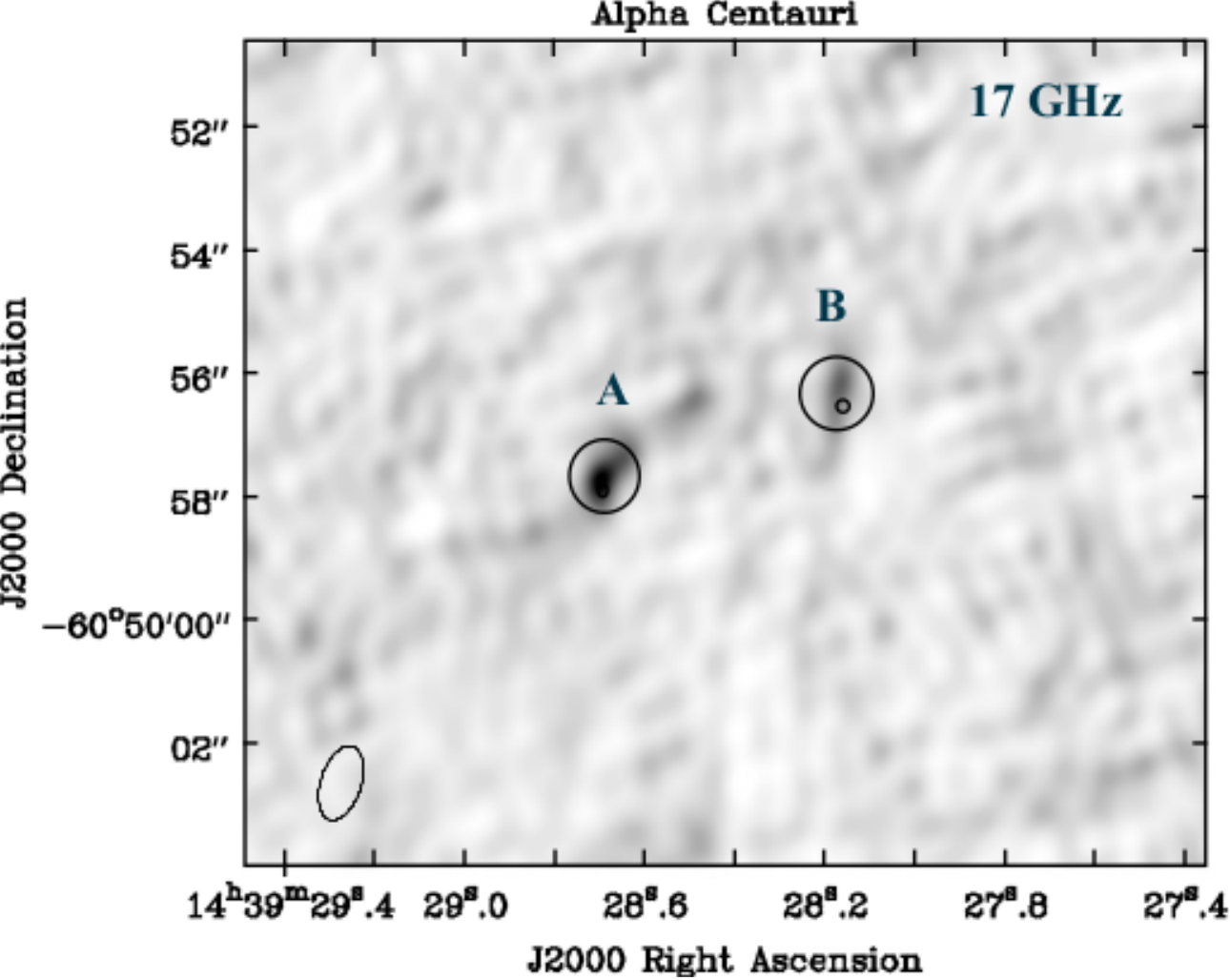}
\caption{Map of the field around \acen at 17 GHz. The two circles labeled as A and B indicate the two stellar components. The two smaller circles indicate the positions computed at the date of the observation (see text).}
\label{mappa-K1}
\end{figure}

\begin{figure}
\includegraphics[width=8.5cm]{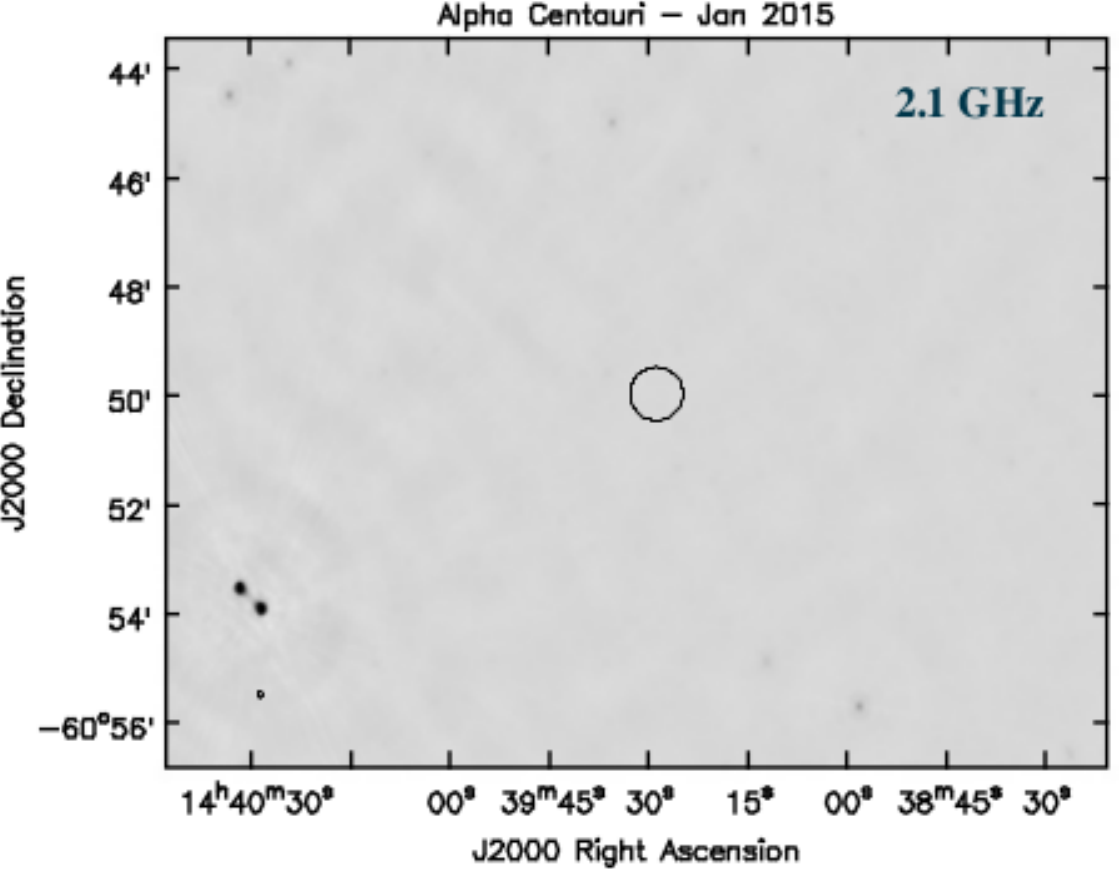}
\caption{Map of the field around \acen at 2.1 GHz after self-calibration applied to the strong double source on the low-left. The circle indicates the position of the two stars. The stars are not detected, only an upper limit due to the noise is given.}
\label{mappa-L}
\end{figure}

\begin{table}
\caption{FK5 Coordinates of $\alpha$ Centauri, Epoch 2015.06}
\label{tab-coord}
\begin{center}
\begin{tabular}{lcc}
\hline\hline
                              &{\acen A}    & {\acen B}   \\
\hline
Computed             & 14:39:28.689  --60:49:57.90  & 14:39:28.158  --60:49:56.53 \\
Observed              & 14:39:28.686  --60:49:57.66  & 14:39:28.171  --60:49:56.31 \\
\hline
\hline
\end{tabular}
\end{center}
\end{table}

\begin{table}
\caption{Measured radio emission of \acen}
\begin{center}
\begin{tabular}{lcccc}
\hline\hline
		&	\multicolumn{2}{c}{17 GHz}			&	\multicolumn{2}{c}{2.1 GHz}		\\
  		& 	Flux density 		& T$_\mathrm{B}$	& 	Flux density	& T$_\mathrm{B}$	\\ 
		&	($\mic{Jy}$)		&		(K)		&	($\mic{Jy}$)	&	(K)			\\		
\hline
\acen A 	&	$161 \pm17$ 		& $13600 \pm1200$  & 	$< 21$ 		&	$< 116\,000$	\\
\acen B 	&	$105 \pm16$ 		& $17800 \pm 2500$ & 	$< 21$ 		&	 $< 230\,000$	\\
\hline\hline
\end{tabular}
\end{center}
\label{dati}
\end{table}

In Fig.\ref{mappa-K1} the field around \acen A and B at 17 GHz is shown. Both stars are clearly detected, with a S/N of 9.5 and 6 respectively see Table~\ref{dati}). The two stars are detected at the centre of the circles centered at the computed positions. Computed and observed coordinates are reported in Table\,\ref{tab-coord}. The synthesized beam is $1.25^{\prime\prime} \times 0.66^{\prime\prime}$, PA=17\deg, allowing us to separate the two stars. The flux densities have been obtained with a two-dimensional gaussian fit of the brightness distribution centred at the star position (task \texttt{IMFIT}). The measured rms of the map around the targets is $15\,\mic{Jy \,beam^{-1}}$. The error in the determination of the flux density has been computed taking into account a 5\% of calibration error. 

At 23 GHz the data were too noisy, due to the high values of the $T_\mathrm{sys}$ and poor phase stability. No valuable images have been produced at this frequency.

At 2.1 GHz, the field around \acen is crowded and many strong sources are present inside the field of view of the main beam of the telescopes (FoV$\approx 20^\prime$). The majority of them are extragalactic sources. In addition, being \acen in the galactic plane ($l=315.7330$, $b=-0.6809$) many galactic sources are also present. Since the main emission process of the extragalactic sources is synchrotron, they can be very bright at low frequency and, therefore, can badly affect the deconvolution process of the maps. The main issues affecting the low frequencies radio observations performed close to the galactic plane have been discussed by \citet{umana2015a}. In particular, there is a strong double source, about 8$^\prime$ far from the phase centre, with a total flux density of about 300 mJy that contaminates the whole field with artifacts. We imaged with the CASA clean task, using multi-frequency synthesis with 3 Taylor coefficients \citep{rau2011}, performing a single round of amplitude and phase self-calibration. After self-calibration the rms of the map at its centre reached $7\,\mic{Jy \,beam^{-1}}$, about an order of magnitude better, approaching the theoretical thermal noise (Fig.~\ref{mappa-L}). The synthesized beam is $6.3^{\prime\prime} \times 5.0^{\prime\prime}$, PA=-27\deg, wider than the separation of the two stars that, therefore, can not be distinguished. However, no source is detected at the positions of \acen, and only an upper limit for their flux density can be set at 3$\,\sigma$ for both stars (Table~\ref{dati}).

We have considered the possibility that the two detections at 17\,GHz are not the two components of A and B, but instead extragalactic sources. However we can exclude this. In fact, if they were extragalactic sources, assuming an average spectral index of -0.7, their flux at 2.1\,GHz should be 0.7 and 0.4 mJy respectively, and they should be detected with high S/N. But no source is seen at 2.1\,GHz.

We also imaged \acen in Stokes V in both bands. No sources have been found in the field. At 17 GHz, we reach an RMS of $13\,\mic{Jy \,beam^{-1}}$, implying that the percentage of circular polarisation is $\pi_\mathrm{c}\mathrm{(A)}\le 8\%$ and $\pi_\mathrm{c}\mathrm{(B)}\le 12\%$ respectively for the two stars. Stokes V maps have the same RMS as the Stokes I ones and, since the system was not detected, we can not provide any upper limit on the percentage of circular polarisation. For a deeper analysis of the circular polarisation of \acen B we refer to Sect.\,\ref{auroral}.

The brightness temperature $T_\mathrm{B}$ was computed from the relation 
\begin{equation}
\label{TB}
T_\mathrm{B}=1960\,\frac{\lambda_\mathrm{cm}^2S_\nu}{\theta^{\prime\prime 2}} \mathrm{(K)}
\end{equation}
with $S_\nu$ the flux density in Jy and $\theta^{\prime\prime}$ the angular diameter of the source in arcseconds, valid in the Rayleigh-Jeans regime when the source is a disk with uniform brightness. Like the case of the Sun, we assume that the radio emission originates few thousand kilometres above the stellar photosphere. Therefore we assume that the radius of the radio source is equal to photospheric radius of the star in the visible and $\theta^{\prime\prime}$ is set 8.502 and 5.999\,mas, as measured with VLTI by \citet{Kervella2017}, for \acen A and B respectively (Table~\ref{param}).

The system \acen was observed by \citet{liseau2016} with ALMA at a difference of few days. There is a clear coincidence in the coordinates observed with ATCA and ALMA and with the procedure by \citet{Phillips}.

\begin{figure*}
\includegraphics[width=8.5cm]{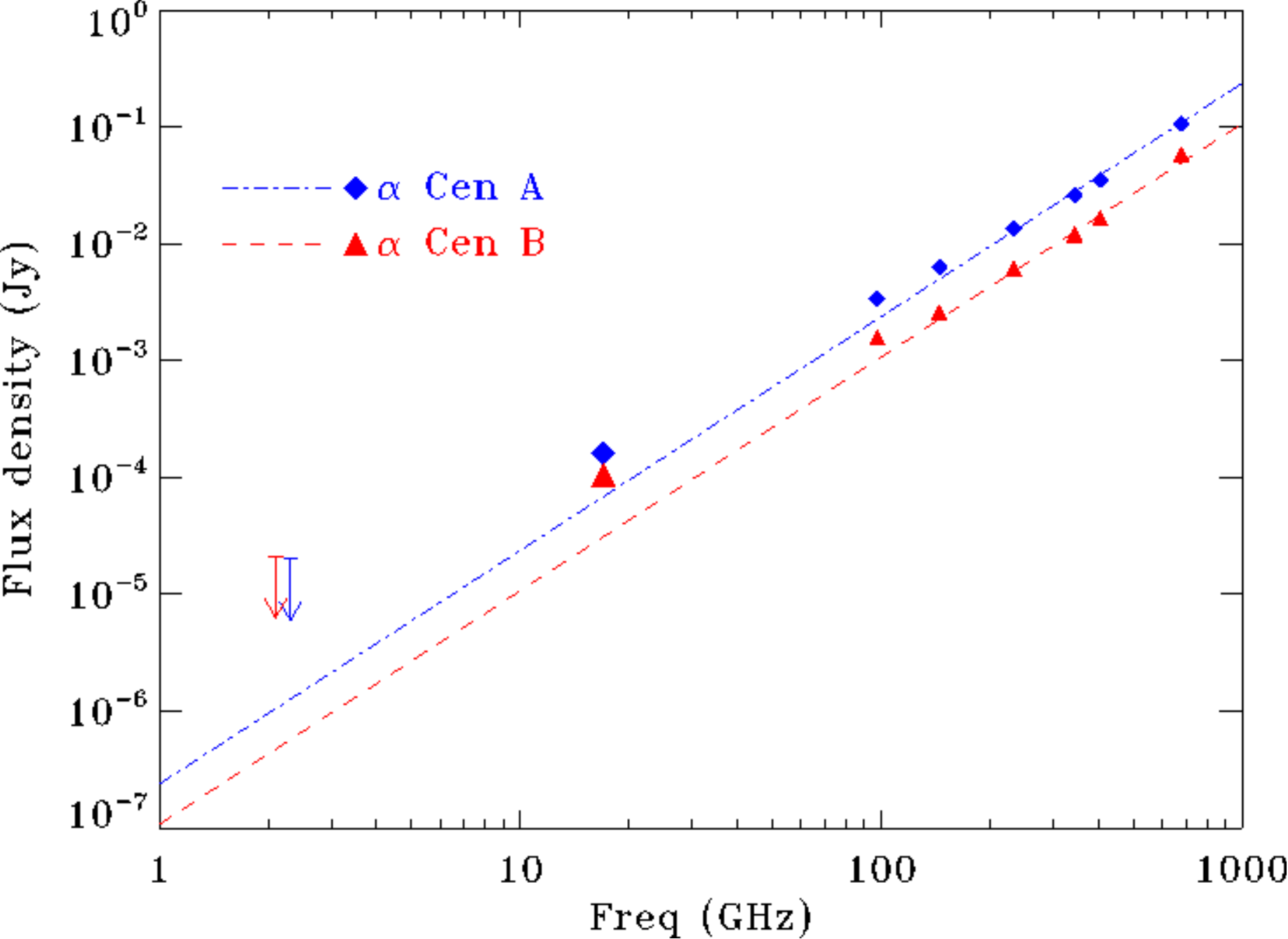}\includegraphics[width=8.5cm]{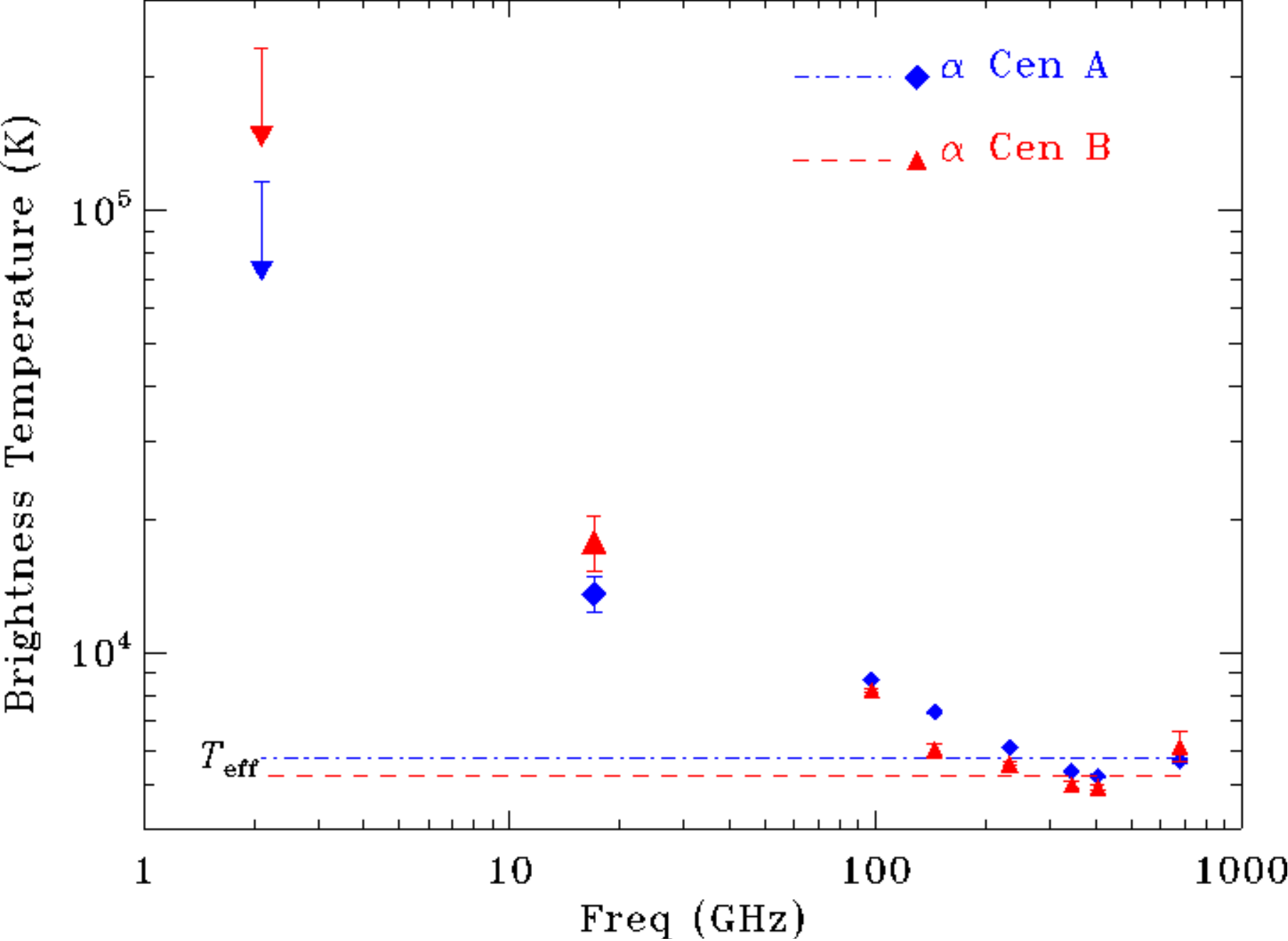}
\caption{Left panel: Flux densities of \acen A and B with ATCA and ALMA; the two dashed lines correspond to black bodies with temperatures of 5824 and 5223 K and same size as the two stars. The two upper limit arrows at 2.1 GHz are at slightly different frequency for a clear visualization. Right panel: with the same notation, the brightness temperature is plotted as a function of the frequency. }\label{fluxes}
\end{figure*}

\section{Discussion}

\subsection{Chromospheric and coronal emission}
Both \acen\ A and B show the same level of chromospheric activity of the Sun \citep{Mamajek2008} as can be deduced by the ratio of the H\&K CaII lines emission to the bolometric one ($R^{\prime}_{\rm{HK}}=5.00$ and $4.92$), as well as the similar rotational periods ($P_{\rm{rot}}=25.6$ and $36.9$ days). Solar values are $R^{\prime}_{\rm{HK}}=4.906$ and $P_{\rm{rot}}=25.05$ days at equator. A study of the X-ray coronal emission of \acen A and B with XMM and Chandra has been published by \cite{ayres2014}, who reports the long term light curves from 2002 to end of 2013. More recently, \cite{Robrade2016} add data up to 2016, including the period of radio observation of this paper. The level of X-ray emission is in general higher for \acen B. At the moment of the radio observations here reported (early 2015) their X-ray luminosity were $L_\mathrm{X}\approx 3\times 10^{26}$ and $10^{27}~\mathrm{erg\,s^{-1}}$ for \acen A and B respectively, comparable to the Sun ($L_\mathrm{X}\approx 2\times 10^{26}$ and $3\times 10^{27}~\mathrm{erg\,s^{-1}}$ for minimum and maximum of activity). The cyclic behavior shows that the X-ray emission is increasing in \acen A and almost at the minimum in \acen B.

The emission originates at optical depth where $\tau\approx 1$; at radio wavelengths probably arises from upper chromosphere, transition region or low corona, while in the millimetric and sub-millimetric range from low chromosphere or photosphere. We collect all the data available at the present. In Table~\ref{tab-emission} we report flux densities measured with ALMA by \citet{liseau2015} together with our ATCA data. $T_\mathrm{B}$ and relative errors have been computed by using eq.~\ref{TB}. In Fig.~\ref{fluxes} (left panel) we report the Spectral Energy Distribution (SED) of \acen A and B in the radio to sub-mm window, together with the $T_\mathrm{B}$ plot. Two blackbody curves corresponding to the $T_\mathrm{eff}$ of the two stars are also reported. From the $T_\mathrm{B}$ plot (right panel), it is evident that there is a minimum at 343.5 GHz, corresponding to the temperature minimum. Then $T_\mathrm{B}$  increases as the frequency decreases reaching values greater than $10^4$K, typical of the top of the solar chromosphere. At 2.1 GHz we have only an upper limit of the flux density, leading to a value of $T_\mathrm{B}<10^5$K, typical of the low corona. It is evident the importance of the radio observations in probing the stellar atmosphere.

\subsection{Comparison to solar chromospheric models}
Observations of free-free emission in the sub-mm, mm and radio bands probe the stellar atmosphere by enabling us to estimate both temperature and density in the emitting region.  We can then compare those quantities to solar chromosphere models.  We estimate density by considering the free-free optical depth for thermal bremsstrahlung. Several models of the atmosphere of the quiet Sun are available in the literature, for different levels of the magnetic activity.  \citet{Vernazza1981}, hereafter VAL, determine chromospheric models for six components, with increasing activity levels (models VAL A to F). Other authors derived more models, but for the purpose of this work we use only models VAL A,B and F, relative respectively to the less bright regions in the solar atmosphere (dark points within granulation cells), quiescent regions (average cell centre) and very bright network elements. For \acen A and B we have only the disk integrated emission, since individual surface features can not be resolved. In the following we use VAL A, B and F in order to represent low, medium and high activity state of the whole stars.

In the case of thermal bremsstrahlung emission, the optical depth at the distance $s$ from the observer, supposed at $s=0$, is given by
\begin{equation}
\label{eq_tau}
\tau_\nu(s)=\int_0^s k_\nu ds
\end{equation}
with the absorption coefficient $k_\nu$ given by
\begin{displaymath}
k_\nu=0.018\, T_\mathrm{e}^{-1.5} n_\mathrm{e}^2 \nu^{-2} g_\mathrm{ff}~~\mathrm{cm}^{-1}.
\end{displaymath}
where $n_\mathrm{e}$ is the plasma number density in $\mathrm{cm}^{-3}$, $T_\mathrm{e}$ the plasma temperature in K and $g_\mathrm{ff}$ the thermally averaged free-free Gaunt factor. In our computation we derived $g_\mathrm{ff}$ by interpolating the data by \citet{vanHoof2014}, that give accurate free-free Gaunt factors in the non relativistic case. We do not use other approximations since they may differ up to more than 20\% from the correct value in the range of $T$ and $\nu$ considered in our analysis.

For \acen A and B it is not possible to directly obtain the profile of $T_\mathrm{B}$ and $n_\mathrm{e}$ as functions of the heigh $h$ above the photosphere. However, assuming that the radiation arises from an atmospheric layer of thickness $\Delta h$, where $\tau_\mathrm{\nu}\approx 1$, we obtain the following relation linking the average $<T_\mathrm{B}>$ and $<n_\mathrm{e}>$:
\begin{equation}
\label{enne}
<n_\mathrm{e}>=\Big(0.018\, <T_\mathrm{B}>^{-3/2}\nu^{-2} g_\mathrm{ff}<\Delta h>\Big)^{-1/2}~~\mathrm{cm}^{-3}
\end{equation}
The comparison of the $<n_\mathrm{e}>$ vs $<T_\mathrm{B}>$ relationship from eq.~\ref{enne} for the three models of the solar atmosphere and \acen A and B will be evaluated in the following.

\begin{figure}
\includegraphics[width=8.5cm]{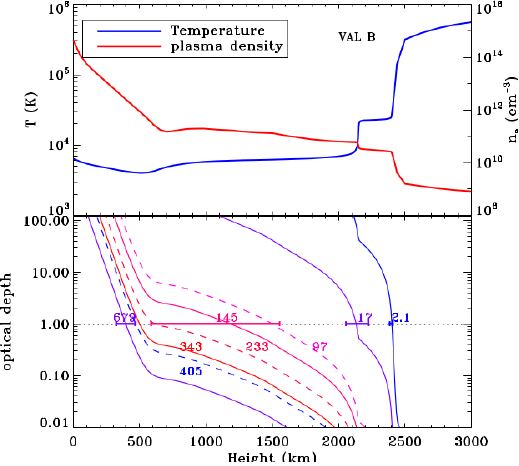}
\caption{Upper panel: Profile of the plasma temperature as a function of the height above the solar photosphere for model VAL B. 
Lower panel: Optical depth for the eight frequencies of the observations of \acen; the thickness of the layers where the continuum radiation arises is indicated by the horizontal segments for four representative frequencies.}
\label{profile}
\end{figure}
\begin{figure}
\includegraphics[width=8.5cm]{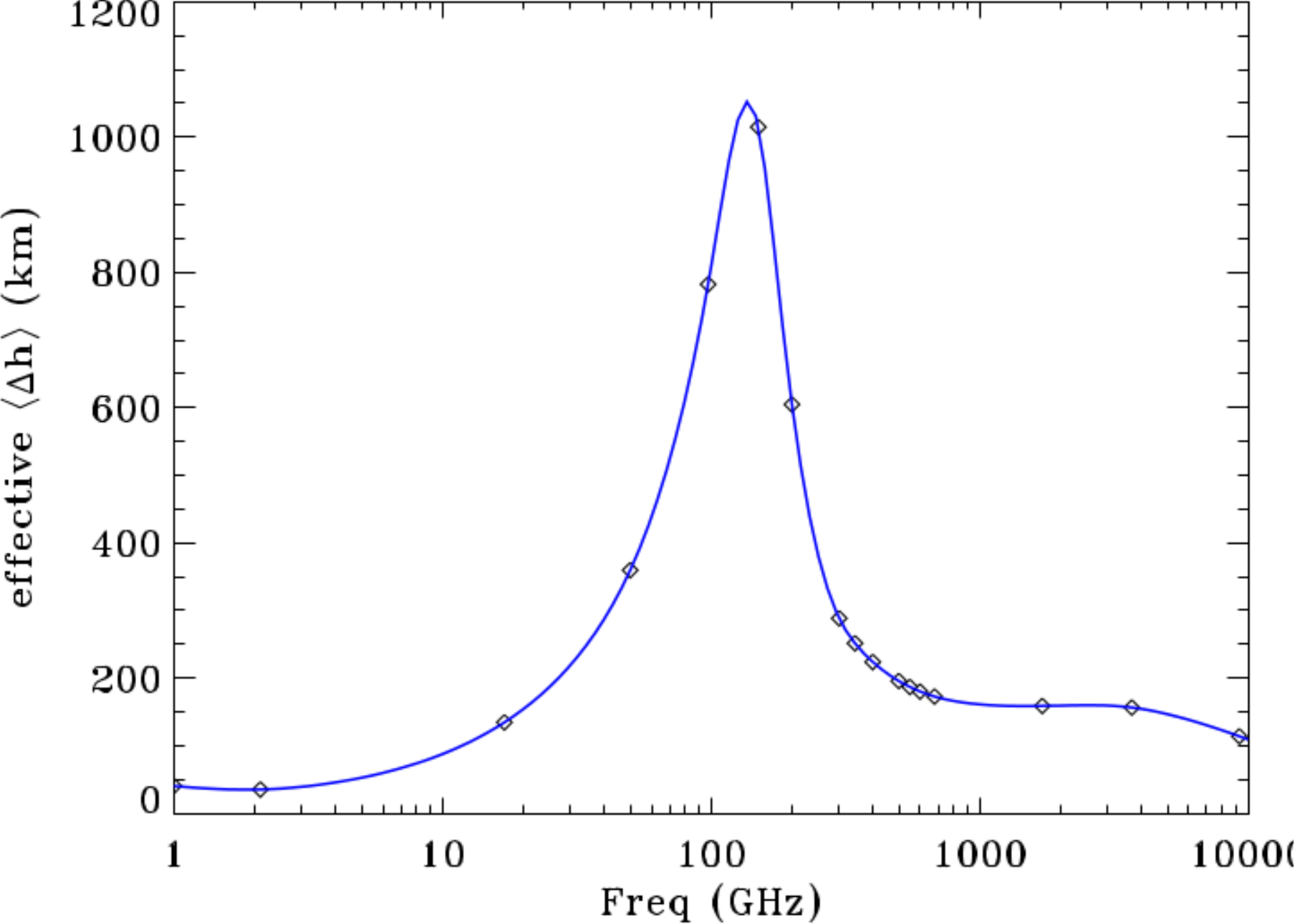}
\caption{Effective depth of the atmospheric layer where the continuum forms, computed for model VAL B. The continuous line is the interpolation of the points (diamonds) where $\Delta$h has been computed numerically.}
\label{delta_h}
\end{figure}

The brightness temperature of emerging radiation, obtained by integrating the transfer equation for a thermal medium in the Rayleigh-Jeans regime, is given by
\begin{equation}
\label{eq_TB}
T_\mathrm{B}=\int_0^\infty T_\nu e^{-\tau_\nu}d\tau_\nu
\end{equation}
where $T_\nu$ is the true temperature at optical depth $\tau_\nu$. $T_\mathrm{B}$ can be computed providing an atmospheric model for which both temperature and plasma density are known.
Temperature and plasma density profiles for VAL B are shown in the top panel of Fig. \ref{profile}.

In order to obtain the average plasma density $<n_\mathrm{e}>$, the thickness $<\Delta h>$ shall be evaluated. For \acen A and B, stellar data do not give the possibility to obtain height and the thickness of the emitting regions. This can be done purely from solar models. We therefore use models VAL A, B and F, that provide tabulated values of $T_\mathrm{e}$ and $n_\mathrm{e}$ as a function of the height $h$ above the solar photosphere at irregular steps $dh$. We first interpolated the model in $N$ regular steps $d h=50$\,km, then computed the optical depth $\tau^{i}_\nu$ from eq.~\ref{eq_tau} for each height $h^{i}$ above the photosphere, as
\begin{equation}
\label{eq_tau_i}
\tau_\nu^{i}=\sum_{j=1}^{i} \Delta \tau_\nu^{j}=\sum_{j=1}^{i} k_\nu(T_\mathrm{e}^{j},n_\mathrm{e}^{j})dh
\end{equation}
with the index $i$ increasing from the outside to the inside. The profiles of $T_\mathrm{e}$ and $\tau_\nu$ for all the observing frequencies (ATCA and ALMA), derived for model VAL B, integrating along the line of sight pointing toward the centre of the star, are shown in the lower panel of Fig.~\ref{profile}.
For each frequency, the brightness temperature is therefore given, integrating eq.~\ref{eq_TB}, by 
\begin{equation}
\label{eq_TB_i}
T_\mathrm{B}=\sum_{j=1}^{N}\Delta\,T_\mathrm{B}^{j}\times e^{-\tau_\nu^{j-1}}
\end{equation}
where
\begin{equation}
\label{eq_delta_T}
\Delta\,T_\mathrm{B}^{j}=T_\mathrm{e}^{j} \times\Big(1-e^{-\Delta\tau_\nu^{j}}\Big)
\end{equation}
is the contribution of each layer $dh$ to the total brightness temperature. The representative height of the layer where the continuum radiation forms is given by the height $h$ where the maximum of $\Delta\,T_\mathrm{B}$ occurs, and its effective thickness by 
\begin{equation}
\label{eq_Dh}
<\Delta h>= \sum_{j=1}^{i}\Delta T_\mathrm{B}^{j}dh /\max (\Delta T_\mathrm{B})
\end{equation}
We show the results of this computation for model VAL B in Fig. \ref{profile}, where $<\Delta h>$ is shown only for 2.1, 17, 145 and 679\,GHz. Here, the horizontal bars are centered at the height $h$ where $T_\mathrm{B}$ is maximum and are limited at the points where $T_\mathrm{B}$ drop to 20\% of the maximum. For the other models, VAL A to F, $<\Delta h>$ does not vary significantly with the frequency, therefore we assume that the results obtained for VAL B are valid for all the state of activity of the stars.

The behavior of $<\Delta h>$ is more evident from Fig.~\ref{delta_h}, where $<\Delta h>$ is computed for several frequencies ad interpolated with a {\it spline} line. Note that $<\Delta h>$ is only about 50 km for 2.1~GHz, then increases reaching a maximum of about 1000 km at 97.5~GHz, and decreasing again for higher frequencies, down to 150 km. This is due to the fact that at 2.1\,GHz the radiation forms in the transition region, where the plasma temperature drops going inwards and, therefore, the absorption coefficient $k_\nu$ increases abruptly. At intermediate frequencies, both plasma temperature and density do not vary too much, and $\tau_\nu$ increases slowly. At higher frequency the continuum forms in the photosphere, after the temperature minimum, where the plasma density increases faster, giving a faster increase of $k_\nu$ and therefore the layer where $\tau_\nu\approx 1$ is quite narrow. 

The derivation of $<n_\mathrm{e}>$ from eq.~\ref{enne} is now possible. 
In Fig.~\ref{model_enne} and Table \ref{tab-emission} we report the derived values of plasma density and brightness temperature for the two stars, including the upper limit at 2.1 GHz, as diamond symbols. 
\begin{table}
\caption{Flux densities and brightness temperatures of \acen in the radio to sub-mm range. $T_\mathrm{B}$ are derived from eq.~\ref{TB} and $<n_\mathrm{e}>$ from eq.~\ref{enne}, using the values of $<\Delta h>$ derived for solar VAL B model.}
\label{tab-emission}
\begin{center}
\begin{tabular}{rcccr}
\hline\hline
Freq     &   Flux density & ref & $T_\mathrm{B}$ & $\log \frac{<n_\mathrm{e}>}{\mathrm{cm}^{-3}}$ \\
(GHz)   &   (mJy)           &     & (K)      &   \\
\hline
\multicolumn{3}{l}{\acen A}\\
     2.1    &  \multicolumn{1}{c}{$<$0.021}     & (1)       &     \multicolumn{1}{c}{$<116\,000$} & $<10.32$ \\  
      17    &  0.161  $\pm$   0.015  & (1) &   13\,600   $\pm$    1\,300 & 10.52 \\
      97    &  3.373  $\pm$   0.011  & (2) &     8\,655   $\pm$          30 & 10.66 \\
    145    &  6.330  $\pm$   0.080  & (3) &     7\,345   $\pm$          93 & 10.73 \\
    233    &  13.58  $\pm$   0.080  & (3) &     6\,100   $\pm$          36 & 11.16 \\
    343    &  26.06  $\pm$    0.19   & (2) &     5\,387   $\pm$          40 & 11.42 \\
    405    &  35.32  $\pm$    0.21   & (3) &     5\,250   $\pm$          31 & 11.50 \\
    679    &  107.2  $\pm$     1.5    & (2) &     5\,670   $\pm$          80 & 11.73 \\
\hline
\multicolumn{3}{l}{\acen B}\\
     2.1    &  \multicolumn{1}{c}{$<$0.021}   & (1) &  \multicolumn{1}{c}{$<230\,000$} & $<10.49$\\
      17    &  0.105  $\pm$   0.015  & (1) &   17\,900   $\pm$    2\,500 & 10.38 \\
      97    &  1.585  $\pm$   0.016  & (2) &     8\,180   $\pm$          82 & 10.57 \\
    145    &   2.580 $\pm$    0.08   & (3) &     6\,020   $\pm$        186 & 10.63 \\
    233    &   6.190 $\pm$    0.05   & (3) &     5\,590   $\pm$          45 & 11.05 \\
    343    &  12.04  $\pm$     0.23  & (2) &     5\,010   $\pm$          95 & 11.30 \\
    405    &  16.53  $\pm$     0.19  & (3) &     4\,944   $\pm$          57 & 11.40 \\
    679    &   57.6   $\pm$      4.5   & (2) &     6\,130   $\pm$        478 & 11.75 \\
\hline\hline
\end{tabular}
\end{center}
references:\\
(1) this paper\\
(2) \citet{liseau2015}\\
(3) \citet{liseau2016}
\end{table}

\begin{table} 
\caption{Solar model VAL A, B, and F predictions of observable properties.}  
\label{tab-model}
\begin{center}
\begin{tabular}{rrrrrrr}
\hline\hline
	& \multicolumn{2}{c}{A}	&	\multicolumn{2}{c}{B}	&	\multicolumn{2}{c}{F}	\\
 Freq& $<T_\mathrm{B}>$ & $\log \frac{<n_\mathrm{e}>}{\mathrm{cm}^{-3}}$& 
            $<T_\mathrm{B}>$ & $\log \frac{<n_\mathrm{e}>}{\mathrm{cm}^{-3}}$& 
            $<T_\mathrm{B}>$ & $\log \frac{<n_\mathrm{e}>}{\mathrm{cm}^{-3}}$\\
(GHz)& (K)~~~ &  & (K)~~~ &   & (K)~~~ &   \\
\hline
2.1 &     32\,700 &     9.91 &    43\,800 &   10.00 &  129\,000 &   10.32 \\
 17 &     12\,800 &   10.28 &    16\,500 &   10.36 &    28\,700 &   10.52 \\
 97 &       6\,310 &   10.52 &      6\,880 &   10.54 &    10\,300 &   10.66 \\
145 &      5\,840 &   10.62 &      6\,360 &   10.65 &      8\,740 &   10.73 \\
233 &      4\,990 &   11.05 &      5\,490 &   11.08 &       7\,470 &   11.16 \\
343 &      4\,530 &   11.31 &      4\,900 &   11.33 &       6\,800 &   11.42 \\
405 &      4\,410 &   11.40 &      4\,720 &   11.42 &       6\,510 &   11.50 \\
679 &      4\,260 &   11.65 &      4\,430 &   11.66 &       5\,660 &   11.73 \\
\hline\hline
\end{tabular}
\end{center}
\end{table}

\begin{figure*}
\includegraphics[width=8.5cm]{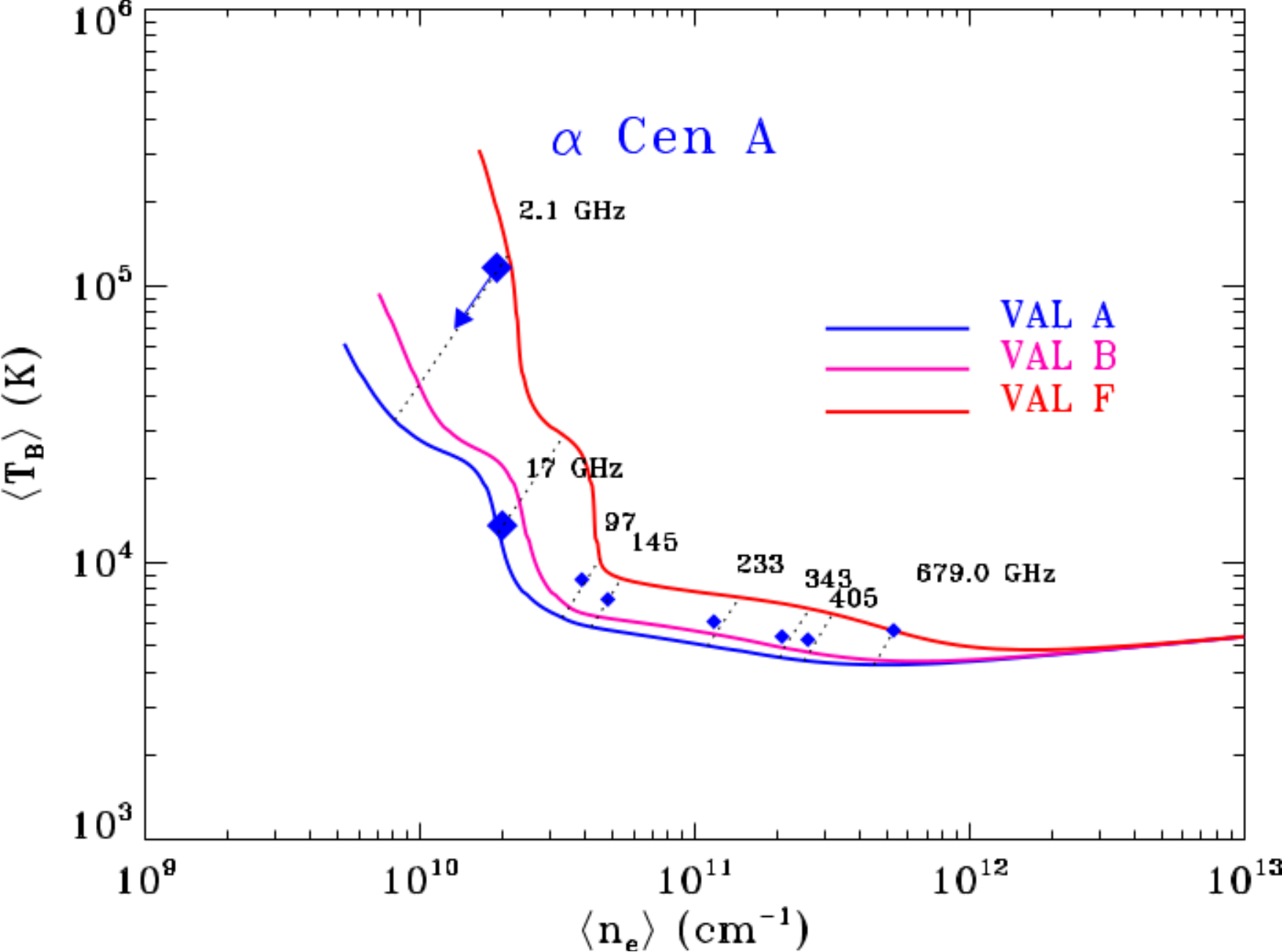}\includegraphics[width=8.5cm]{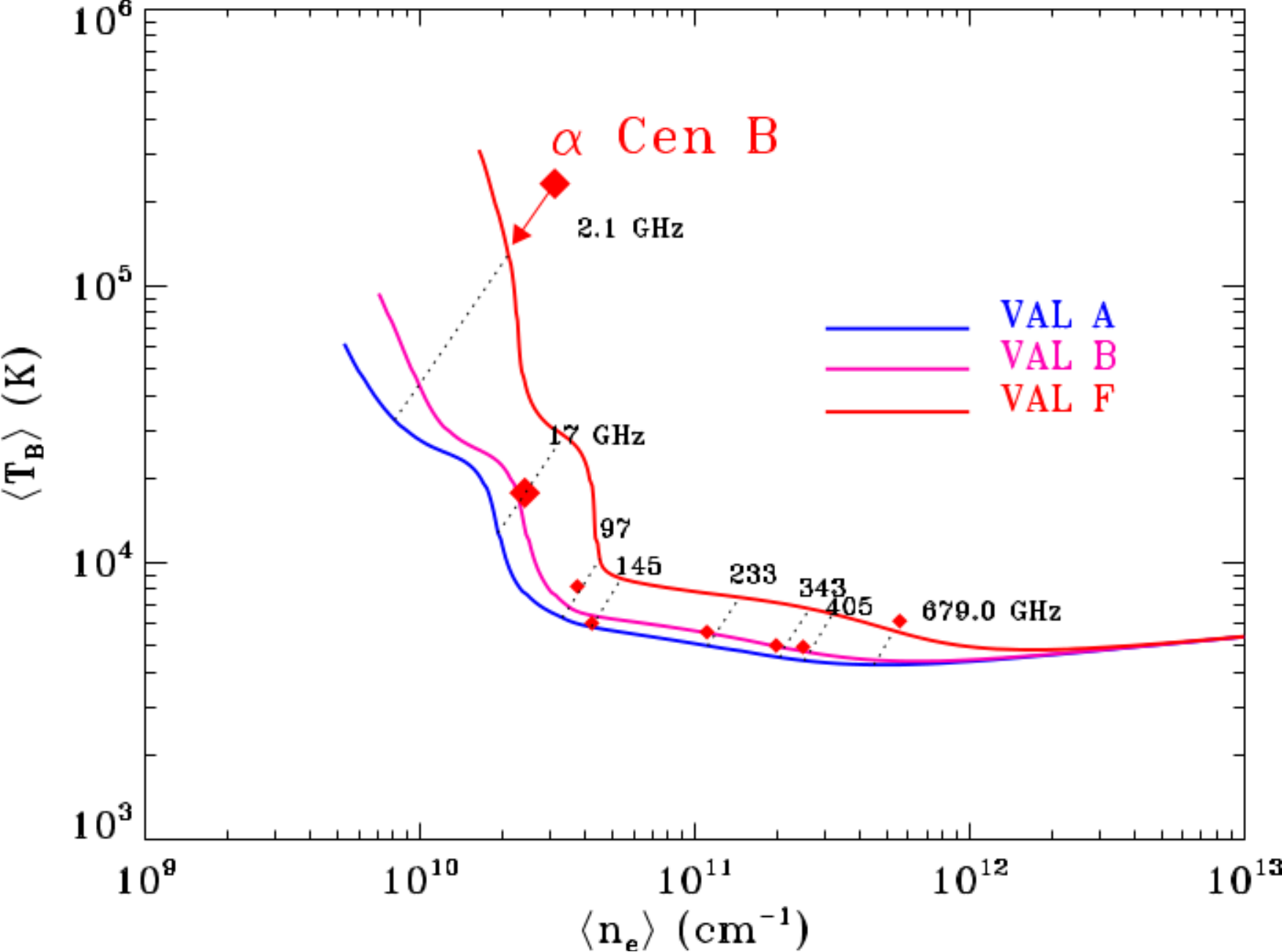}
\caption{Left panel: Average brightness temperature as a function of the average plasma density number for \acen A derived from the observations (big diamonds: ATCA; small diamonds: ALMA) compared to the solar disk-integrated brightness temperature for models VAL A, B and F, shown as continuous lines; dashed lines connect values relative to the observed frequencies. Right panel: same for \acen B. Data at 2.1 GHz are upper limits for both stars and are indicated by arrows.}
\label{model_enne}
\end{figure*}

In order to compare the level of activity of \acen A and B with the Sun, for each of the three models A, B and F, we compute the solar integrated disk brightness temperature $<T_\mathrm{B}>$  and the corresponding number densities $<n_\mathrm{e}>$ of the chromospheric plasma as before. For this purpose, for each frequency in the range 1-10\,000 GHz, we compute $\tau_\nu$ with eq.~\ref{eq_tau_i} along the line of sight for different value of the radii $r$ from the centre of the solar disk, then compute the corresponding $T_\mathrm{B}$ with eq.~\ref{eq_TB_i} and eventually $<T_\mathrm{B}>$. Average number densities have been computed by using eq.~\ref{enne}. Results are shown in (see Table~\ref{tab-model}) and drawn in Fig.~\ref{model_enne} as continuous lines together with the \acen A and B data.

The comparison between the data of \acen A and B and the three models A, B and F indicates that the level of activity of the two stars is almost the same as in the case of the Sun.With the exception of the data at 2.1\,GHz, where we have only upper limits for $T_\mathrm{B}$, the errors in the measured brightness temperatures are much lower than the difference of $T_\mathrm{B}$ for the two contiguous models A and B, indicating that the method described in this paper is adeguate to infer the state of activity of solar-type stars based on the analysis of the free-free emission in the centimeter to sub-millimeter wavelength range. The points at 17 GHz seem to indicate that \acen A is in a low state of activity at the moment of the observations, as it lies close to the model A line, relative to to dark regions of the chromosphere. \acen B seems to show a higher level of activity, as the point at 17 GHz lies slightly above the model B line, relative to the quiescent state. This is consistent with the observed level of emission in the X-rays \citep{Robrade2016} who observed \acen in the same days (2015 Jan 24) of the ATCA observations. ALMA observations at the different bands are not simultaneous each other, and this can cause the fact that the data points do not seem to follow exactly the same model line. In addition, the X-ray data are too far in time and the activity level can not be compared directly.

\subsection{The filling factor of active regions}
\label{filling}

One of the emission mechanisms that can occur in the Sun and solar-type stars is the thermal gyroresonance. In presence of a magnetic field, the emission from a non-relativistic electrons can occur at the $s^\mathrm{th}$ harmonic of the Doppler-shifted gyrofrequency $\nu_\mathrm{B}=2.8\times 10^6 B$\,(Hz), i.e. $\nu\approx s\,\nu_\mathrm{B}$ (see \citealp{dulk1985} for an overview). For a distribution of electrons, the Doppler shift depends on the velocity, and therefore, in a thermal plasma, the overall resonance condition depends on the temperature $T$. The absorption coefficient at the frequency $s \nu_\mathrm{B}$, for extraordinary and ordinary magnetoionic modes, is given by \citep{gudel2002}:
\begin{equation}
\label{kgiro}
k_\nu(s)=1020(1\pm0.5)^2\frac{n_\mathrm{e}}{\nu T^{1/2}} \frac{s^2}{s!} \Big( \frac{s^2 T}{1-6\times10^{10}}\Big)^{(s-1)}\gamma_\nu(s)
\end{equation}
where $\gamma_\nu(s)$ is a gaussian-like profile around the central frequency given by
\begin{equation}
\label{larghezzagiro}
\gamma_\nu(s)=\exp \Big[ -\frac{(1-s \nu_\mathrm{B}/\nu)^2} {8.4\times 10^{-11}T}\Big].
\end{equation}
For a typical magnetic field above a solar active region of the order of 1\,kG and temperature respectively $10^4$, $10^5$ and $10^6$\,K, the width of the $\gamma_\nu$ function is approximately $\Delta\nu_\gamma=4$, $15$ and $40$\,MHz. The thickness $L$ of the region where gyroresonance occurs depends on the vertical gradient of the magnetic field
\[
L=\frac{\Delta\nu_\gamma/2.8\times 10^{6}}{ d B/dz}
\]
In the Sun, typical values of $d B/dz$ range from 0.2 to 2 G\,km$^{-1}$ \citep{solanki2003}. Assuming 1\,G\,km$^{-1}$, the corresponding thickness $L$ is therefore 1.5, 5 and 15 km, giving an optical depth $\tau_\mathrm{B}=k_\nu(s)L$ that ranges around $10^6-10^8$. Concluding, a typical magnetic field of thousand G makes the solar and stellar atmosphere optically thick at the highest level, in the corona. As a consequence, the brightness temperature above the active regions saturates to the local thermodynamical temperature, that in the corona is of the order of $10^6$\,K.

The upper limit of $T_\mathrm{B}^\mathrm{max}$ at 2.1\,GHz (see Table\,\ref{tab-emission}) can be used to derive an upper limit of the filling fraction $f$ of stellar surface covered by active regions. Given $f$, the disk integrated brightness temperature is given by 
\[
T_\mathrm{total}=(1-f)\times T_\mathrm{model} + f\times T_\mathrm{cor}
\]
where $T_\mathrm{model}$ is the theoretical brightness temperature from VAL models (see Table\,\ref{tab-model}) and $T_\mathrm{cor}$ the  coronal one. 

Since $T_\mathrm{total}<<T_\mathrm{B}^\mathrm{max}$, using VAL A model for \acen A (see Fig.\,\ref{model_enne}), we infer $f<<0.08$. For \acen B, using instead VAL B model, $f<<0.2$.

\subsection{The Search for Time-Variable Coherent Radio Emission}
\label{auroral}

Impulsive high circularly polarized coherent radio bursts, generated by the Electron Cyclotron Maser emission process, have been detected at the low frequencies in the Sun and other stars \citep{melrose_dulk82}. The close similarity of the stars forming the \acen system with our Sun makes reasonable to expect stellar coherent bursts also from \acen. Even if the \acen system has not been detected at the frequency of 2.1 GHz, short duration events like ECM pulses may occur and not be seen in a map that average in time all the data. It is not possible to analyze dynamical spectra for each star separately since the angular resolution is too coarse (see Sect.\,\ref{results}). In addition, since the ECM radiation is circularly polarized, we performed an analysis of the data in Stokes V and time. We observed for about 12 hours per day in three days (Table\,\ref{log-obs}), covering less than 50\% of the period. Dynamical spectra in Stokes V have been obtained following the same method as in \citet{trigilio2008}. We computed the discrete Fourier transform (DFT) as a function of the visibilities $V(\nu,t)$ at the position of the source. For each spectral channel at frequency $\nu$ and for each time interval $t$, the flux density $F(\nu,t)$ is given by
\begin{equation}
\label{DFT}
F(\nu,t)=\frac{1}{N}\Sigma_{j=1}^N V_{j}(\nu,t)e^{-2\pi\,i(u_{j}x_0+v_{j}y_0)}
\end{equation}
where $x_0$ and $y_0$ are the offsets of the target source from the phase tracking centre in RA and Dec respectively and $N$ the total number of visibilities. After RFI removal, within a total bandwidth of 1800 MHz, bin of 100 MHz and time resolution of 10 s, no enhancement due to coherent emission is detected in Stokes V.

In case of stars hosting a planetary system, coherent radio bursts could be also triggered by the magnetic interaction between the star and the planets (SPMI). The SPMI could be able to power amplified auroral radio emission in the stellar flux tube crossed by the planet. In the solar system, all the magnetized planets show auroral emission in several bands of the spectrum, including the radio band. The auroral radio emission could be powered by the field-aligned current system that originates in the outer magnetospheric region, where the co-rotation breakdown of ionized material takes place \citep{cowley_bunce01}. But there are also evidence of planetary auroral radio emission triggered by other two possible causes: a) the impact of the stellar wind into the planetary magnetosphere, as in the case of Earth, Jupiter, Saturn, Uranus and Neptune; b) magneto hydro dynamic shocks triggered by magnetized satellites, the dipolar inductors, as in the case Jupiter with Ganymede, or unmagnetized, unipolar inductors, as in the case of Jupiter with Io and Europa (see \citealt{zarka2007} for an extensive review). In all these cases, electrons accelerated to few keV propagate along converging dipolar field lines toward the planet. Part of them precipitate into the atmosphere \citep{badman2015} causing auroral emission (optical and UV lines, X-ray emission), other are reflected back by magnetic mirroring. The lack of reflected electrons with small pitch angles (the ones precipitated into the atmosphere) reflects in an anisotropy in the velocity space that is the cause of the cyclotron maser instability and consequent ECM emission, the auroral radio emission (ARE). The radiation propagates almost perpendicular to the local magnetic field and is 100\% circularly polarized  with very high brightness temperature. 

Several stars, of different stellar type, show similar mechanism for ECM. The first star discovered to show ARE was CU\,Vir \citep{trigilio2000}, a magnetic A0V star surrounded by a large magnetosphere shaped like a simple dipole, that exhibits two pulses during its rotational period at 1.4 and 2.2\,GHz, observed over more than a decade \citep{trigilio2008, trigilio2011}. In the centimeter wavelengths range ARE has been detected in some ultra cool MS dwarfs (UCDs, spectral type higher than M7). The overall description of the ARE from the UCDs and an updated list of UCDs showing ARE are given by \citet{pineda_etal17}.

Following the approach developed by \citet{hess2011} to simulate the ARE visibility, some periodic features of the ARE from the M8.5 dwarf TVLM513-46 \citep{hallinan2007} have been explained by \citet{leto2017} assuming that the same process occurring in the interaction between Jupiter and its satellites could occur also between star and planet (SPMI). Fig.\,\ref{ECME2} shows a schematic view of this scenario. 

The amplification mechanism responsible of the  stellar ARE is the ECM. This mechanism amplifies  the first or second harmonic of the local gyrofrequency,  $\nu_\mathrm{B}$. Therefore the frequency range that could be amplified by the ECM is proportional to the magnetic field of the star, much more intense than the planetary one. For Jupiter $B_\mathrm{pole}\approx 10$\,G, for CU\,Vir $3000$\,G, for UCDs is of the order of kG \citep{Reiners2010}. In the case of a KV star, like \acen B, the overall topology of the magnetosphere is not dipolar, since more complex magnetic structures characterize the magnetic activity and stellar cycles. However, a  fraction $f$ of the photosphere is covered by spots where the magnetic fields  is of the order of of kG \citep{Saar2001}. For \acen B we have determined an upper limit for $f$ (see Sect.\,\ref{filling}). If ARE is powered along the field lines anchored with such magnetic spots, the ECM emission could be detected in the GHz frequency range. However, the occurrence of the ARE depends on the topology of the magnetosphere of the star, that can be dipolar, toroidal and even variable, the actual frequency on the magnetic field strength at the surface, the visibility from the Earth on directivity of the emission.

\begin{figure}
\includegraphics[width=8.5cm]{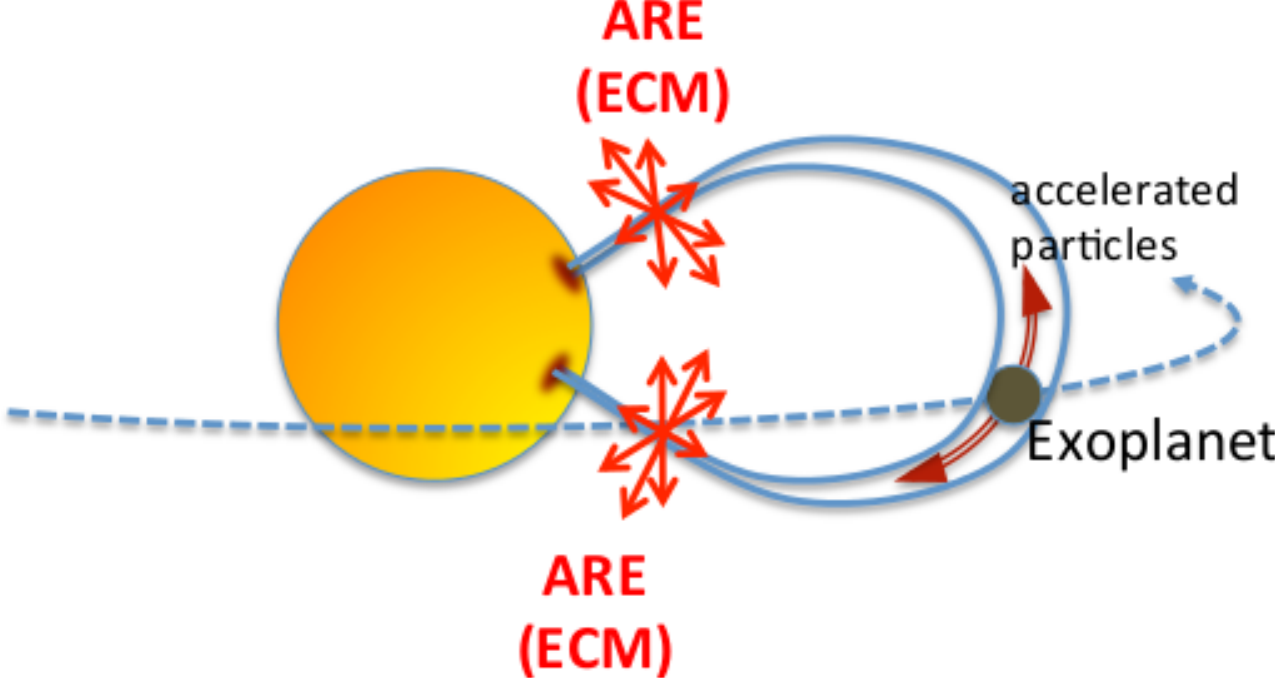}
\caption{Schematic view of the auroral radio emission due to star-planet magnetic interaction. Particle acceleration is triggered by the planet passing through a magnetic loop of the star. The circular polarization of the ECM emission has opposite helicity in the North and South hemispheres.}
\label{ECME2}
\end{figure}

The analysis of the Stokes V dynamical spectra of  \acen could be also useful to search coherent bursts due to SPMI. In the case of a planet orbiting close to the star exists, the ARE will be seen as periodic phenomenon. Although the recent claimed discovery of a planet, \acen\ Bb, supposed in close orbit with its parent star \citep{dumusque2012}, is doubtful \citep{Rajpaul2016},  the detection of periodic coherent burst could be signature of SPMI. Looking at the Stokes V dynamical spectrum of \acen performed in the 1--3 GHz frequency band, there is no suggestion of ARE. After averaging the dynamical spectra in 2 minutes intervals over the whole bandwidth, no evidence of bursts at Stokes V, neither positive nor negative, was found with a $3\sigma$ confidence level of $0.5\,\mathrm{mJy}$. Possible interpretations for this are:
\begin{itemize}
	\item[-] the ECM occurs at this frequency but, since it is highly beamed, it is never directed toward Earth for the geometry of the system with respect to the line of sight;
	\item[-] the ECM occurs at this frequency but the flux density is too low for the sensitivity of the observations;
	\item[-] the coverage of the orbital period of the hypothetical planet (3.236 days) is only $\approx 50\, \%$ and, even if present, the ECM is not direct toward Earth in the periods covered by the observations;
	\item[-] the ECM occurs at another frequency;
	\item[-] there is no magnetic interaction between planet and parent star;
	\item[-] there is no planet close to \acen\ B, consistent with the findings of \cite{Rajpaul2016}.
\end{itemize}

\section{Conclusions}
The new generation of radio interferometers have reached a sensitivity that allowed us to detect, for the first time, the quiet centimeter-wavelength emission from solar-type stars \acen A and B. In fact, despite the proximity to the Earth, the radio emission of \acen A and B is very faint at centimetric wavelengths. We detected both stars at 17 GHz and an upper limit is set at 2.1 GHz. 

The measured flux densities at radio wavelengths, together with the measurements with ALMA in the sub-millimeter, allow us a comparison with the solar atmosphere. Based on the model of the solar atmosphere available in the literature, we developed a method to relate the average temperature and plasma density in the layers where the continuum radiation forms. For \acen A and B the atmospheric structure seems very similar to the solar one, as expected. It is important to note that the deep radio observations that will be carried out with the next generation of radio interferometers, as the Square Kilometer Array (SKA) will give the opportunity to study the structure of the upper chromosphere and low corona of solar-type stars within tens of pc from the Sun \citep{umana2015b}.

In response to the claimed detection of a planet in a close orbit around \acen B, even if doubtful, we searched for auroral radio emission from the star triggered by the planet passing through the stellar magnetosphere, assuming a mechanism similar to that occurring between Jupiter and Io. Spectro-temporal analysis of the data shows no cyclotron maser in the band 1-3 GHz, at least with a limiting sensitivity of 0.5 mJy. 


\end{document}